\documentclass[prd,twocolumn,showpacs,preprintnumbers,nofootinbib,amsmath,amssymb,nobalancelastpage]{revtex4}
\usepackage[pdftex]{graphicx}
\usepackage{latexsym,amsmath,amssymb,lmodern,url}
\usepackage{natbib}
\usepackage[pdftex,bookmarks,linktocpage,pdfpagelabels,plainpages=false,hyperfigures,linkcolor=blue,citecolor=blue,urlcolor=blue]{hyperref} \usepackage{xspace}
\usepackage{courier}
\hypersetup{colorlinks=true}
\usepackage{bm}
\usepackage{dcolumn}
\usepackage{amsmath}
\usepackage{amssymb}
\usepackage{color}
\usepackage{xcolor}
\usepackage{soul}
\usepackage{url}

\usepackage{aas_macros}
\usepackage[normalem]{ulem}

\graphicspath{{./figs/}}

\newcolumntype{L}{>{\centering\arraybackslash}m{10cm}}

\def\rhosoliton{{\rho_{\mathrm{soliton}}}}

\def\Mvir{{M_\mathrm{vir}}}
\def\rhoNFW{\rho_{\mathrm{NFW}}}
\def\rhoGM{\rho_{\mathrm{GM}}}
\def\rhosol{\rho_{\mathrm{sol}}}
\def\rsol{r_{\mathrm{sol}}}
\def\M200{M_{200}}
\def\c200{c_{200}}
\def\m22{m_{22}}
\def\Msun{M_\odot}
\def\MMW{M_{\mathrm{MW}}}
\def\MBH{M_\mathrm{BH}}

\begin{document}

\title{The viability of ultralight bosonic dark matter in dwarf galaxies}

\author{Isabelle S. Goldstein}
\email{isabelle\_goldstein@brown.edu}
\affiliation{ Department of Physics, Brown University, Providence, RI 02912-1843, USA}
\affiliation{ Brown Theoretical Physics Center, Brown University, Providence, RI 02912-1843, USA}

\author{Savvas M. Koushiappas}
\email{koushiappas@brown.edu}
\affiliation{ Department of Physics, Brown University, Providence, RI 02912-1843, USA}
\affiliation{ Brown Theoretical Physics Center, Brown University, Providence, RI 02912-1843, USA}

\author{Matthew G. Walker}
\email{mgwalker@cmu.edu}
\affiliation{McWilliams Center for Cosmology, Department of Physics, Carnegie Mellon University, Pittsburgh, PA 15213}

\date{\today}

\begin{abstract}
The dark matter distribution in dwarf galaxies holds a wealth of information on the fundamental properties and interactions of the dark matter particle. In this paper, we study whether ultralight bosonic dark matter is consistent with the gravitational potential extracted from stellar kinematics. We use velocity dispersion measurements to constrain models for halo mass and particle mass. The posterior likelihood is multimodal. Particle masses of order $m\sim 10^{-22} {\rm{eV}}$ require halos of mass in excess of $\sim 10^{10} \Msun$, while particle mass of order $m \gtrsim 10^{-20}{\rm{eV}}$ are favored by halos of mass $\sim [10^{8} - 10^{9}] \Msun$, with a similar behavior to cold dark matter. Regardless of particle mass, the lower halo masses are allowed if stellar dynamics are influenced by the presence of a central black hole of mass at most $\sim 10^{-2}$ the host halo mass. We find no preference for models that contain a black hole over models that do not contain a black hole. Our main conclusion is that either the fuzzy dark matter particle mass must be $m \gtrsim 10^{-20}$ eV, or the Milky Way dwarfs must be unusually heavy given the expected hierarchical assembly of the Milky Way, or the Milky Way dwarfs must contain a central black hole. We find no evidence for
either of the last two possibilities and consider them unlikely.
\end{abstract}

\maketitle

%%%%%%%%%%%%%%%%%%%%%%%%%%%%%%%%%%%%%%%%%%%%%%%%%%%%%%%%
%            		Intro     
%%%%%%%%%%%%%%%%%%%%%%%%%%%%%%%%%%%%%%%%%%%%%%%%%%%%%%%%

\section{\label{sec:intro} Introduction} 

In the last 20 years, Milky Way dwarf galaxies have provided a wealth of information regarding the nature of dark matter. 
As dark matter dominated systems with mass to light ratios in excess of $M/L \gtrsim 10$, and devoid of most baryonic astrophysical complexities, they are ideal hypothesis testing systems for some of the most fundamental properties of dark matter. Dwarf galaxies have been used to place the most robust to-date constraints on the annihilation \cite{fermiAnnihilation2011, GeringerSameth2011}, decay \cite{GeringerSameth2015, 2015MNRAS.453..849B, 2010JCAP...12..015D}%JCAP paper is galaxy clusters
, and self-interaction cross sections \cite{PhysRevLett.116.041302}. In addition, their mere existence places limits on whether the dark matter particle decoupled while relativistic.\cite{PhysRevLett.42.407}. 

The distribution of dark matter in dwarfs is a subject of debate. Collision-less, cold dark matter gives rise to cusps, while alternative dark matter models, like self-interacting dark matter \cite{2019JCAP...04..026B,2013PDU.....2..139F} and warm dark matter (free streaming from non-zero velocities at decoupling) \cite{2006PhRvD..74b3527A, 2017arXiv171204597W, 2017PhR...711....1A}, predict central profiles that are cored. The formation and dark matter distribution of these objects through complex baryonic galaxy formation arguments has been reproduced in simulations, and the underlying physics has been at the forefront of fundamental dark matter study.

The tool of choice in all the aforementioned studies is reconstructing the gravitational potential using stellar kinematics (e.g., \cite{GeringerSameth2011, PhysRevD.75.083526, 2008ApJ...678..614S}). Measurements of stellar velocities along the line of sight of bright red giants allows for the distribution of dark matter to be reverse engineered \cite{2013pss5.book.1039W}. Such measurements are extremely powerful because under a set of reasonable dynamical assumptions they allow for robust potential reconstruction with well controlled errors, especially in the case of  classical dwarf galaxies. Attempts to apply such methods to the faintest objects in the universe (ultra-faint dwarf galaxies discovered in the last few years) carry much larger uncertainties and thus are less constraining  \cite{GeringerSameth2015,2015MNRAS.453..849B}.

In this paper we use stellar kinematics to test the viability of ultralight bosonic dark matter in dwarf galaxies. Motivations for such dark matter candidates come from GUT-scale physics, originally introduced through the solution to the strong CP problem in quantum chromodynamics \cite{PecceiQuinn1977A, PecceiQuinn1977B, Weinberg1978}, and subsequently envisioned through cosmology and large scale structure \cite{Hui2017, Marsh2016, Suarez2014,PhysRevLett.85.1158,2014MNRAS.437.2652M,2014NatPh..10..496S,PhysRevD.63.063506,Matos_2000,PhysRevD.62.103517,PhysRevLett.64.1084,PhysRevD.50.3650,PhysRevD.64.123528,1985MNRAS.215..575K,2021EPJC...81..286P,1983PhLB..120..133A,1983PhLB..120..137D,1983PhLB..120..127P}. 
Qualitatively, such objects go by the name of ``fuzzy dark matter", a term that denotes a fundamental characteristic property: the existence of a coherent quantum state (a Bose-Einstein condensate), described by the Schr\"odinger-Poisson equation and forming soliton cores instead of cusps \cite{Schive2014,Mocz2018}, for a thorough review please see \citet{Hui2017}. 

The quantum pressure of fuzzy dark matter arises from ultralight bosons mass $\sim 10^{-22}$ eV or scalar field DM with de Broglie wavelength about the size of the dwarf galaxy stellar component ($\sim 1$ kpc).  The existence of a soliton core suppresses small scale structure \cite{Hui2017,Schutz2020} (see \cite{Safarzadeh_2020,aviloeb2022} for additional constraints from Milky Way satellites). Throughout the paper we will be using the terms fuzzy dark matter and/or axion-like dark matter interchangeably to refer to dark matter that forms quantum pressure supported soliton cores. 

We examine the viability of ultralight boson dark matter in dwarf galaxies using stellar kinematics in six classical dwarf galaxies: Fornax, Sculptor, Draco, Sextans, Ursa Minor, and Carina. We choose to use only classical dwarf galaxies because they contain enough stars and observations ($N_\mathrm{obs}\gtrsim 500$) to provide meaningful constraints on the dark matter gravitational potential. 

The primary goal of this work is to determine whether ultralight bosonic dark matter is consistent with velocity dispersion measurements, and if so, what range of parameters allow such consistency. We find that unless the Milky Way did not have a typical evolution, the mass of the ultralight dark matter particle must be at least $m > 10^{-20} {\rm eV}$. 

The paper is organized as follows: In Section II we outline the reconstruction of the mass distribution using stellar velocity dispersion measurements (Jeans analysis). In Section III we review a set of dark matter density profiles that have been proposed in the literature as soliton solutions to dark matter halos. Section IV summarizes the observations used in the paper, and in Section V we present the results. We conclude in Section VI. 

%%%%%%%%%%%%%%%%%%%%%%%%%%%%%%%%%%%%%%%%%%%%%%%%%%%%%%%%
%            			Jeans eqn     
%%%%%%%%%%%%%%%%%%%%%%%%%%%%%%%%%%%%%%%%%%%%%%%%%%%%%%%%

\section{\label{sec:SecII}Stellar kinematics potential tracers}

Line of sight stellar velocity measurements from dwarf galaxies can be used in the construction of a stellar velocity dispersion profile (velocity dispersion as a function of radius from the center of the dwarf). The velocity dispersion traces the underlying matter density distribution. 

The spherical Jeans equation allows for the reconstruction of the gravitational potential, $\Phi(r)$ given a velocity dispersion profile \cite{GalacticDynamics}, 
%	EQUATION
\begin{eqnarray}
\label{eqn:jeans}
\frac{d\Phi}{dr} &=& - \frac{G M(r)}{r^2} \nonumber \\
&=& - \frac{1}{\nu(r) } \frac{d}{dr}  \left[ \nu(r) \overline{u_r^2}(r) \right] - 2 \frac{\beta_a (r)  \overline{u_r^2}(r)}{r}.
\end{eqnarray}
Here, $\beta_a$ the orbital anisotropy is a measure of the difference between tangential and radial dispersions, 
%	EQUATION
\begin{equation}\label{eqn:betaA}
\beta_a(r) \equiv  1 - \frac{2 \overline{u_\theta^2}(r) }{\overline{u_r^2}(r)}  ,
\end{equation}
and $\nu(r)$ is the stellar density profile, with $\overline{u^2_r}(r)$ the radial stellar velocity dispersion profile, 
%	EQUATION
\begin{equation}\label{eqn:dispComponents}
\overline{u^2}(r) = \overline{u^2_r}(r) + \overline{u^2_\theta}(r) + \overline{u^2_\phi}(r). 
\end{equation}
The mass enclosed is $M(r)$ defined in the usual way, 
%	EQUATION
\begin{equation}\label{eqn:mass} 
M(r) = 4\pi \int_0^r s^2 \rho(s) ds   .
\end{equation}
Under the assumption of spherical symmetry and dynamical equilibrium, Equation (\ref{eqn:jeans}) has the general solution:
%
%	EQUATION
\begin{equation}\label{eqn:genSoln} 
\nu(r) \overline{u_r^2}(r) = \frac{1}{f(r)} \int_r^\infty f(s) \nu(s) \frac{G M(s)}{s^2}ds, 
\end{equation}
where $f(r)$ is 
\begin{equation}\label{eqn:functionfr}
f(r) = 2\,f(r_1) \exp\left[ \int_{r_1}^r \beta_a(s) s^{-1}ds\right].
\end{equation} 

The assumption of dynamic equilibrium is implicit in the Jeans equation, but unlikely to hold precisely for all of the Milky Way satellites that we consider.  Nevertheless, various studies have shown that violation of this assumption is unlikely to have dramatic effects on the inferred dynamical mass \cite{1995ApJ...442..142O,1995AJ....109.1071P, 2006MNRAS.367..387R}.  In any case, the dynamical crossing time of a typical Milky Way dwarf spheroidal is a small fraction of its orbital period, such that we can expect a state of near equilibrium to hold over most of the dwarf's orbit.

If we assume the orbital anisotropy is a constant within a given system, then the velocity dispersion projected along the line of sight is 
\begin{equation}\label{eqn:losveldisp}
\sigma^2(R) \Sigma(R) = 2 \int_R^\infty \left(1-\beta_a(r)\frac{R^2}{r^2} \right) \frac{\nu(r)\,\overline{u_r^2}(r)\, r}{\sqrt{r^2-R^2}}dr, 
\end{equation}
where $R$ is the projected radial distance from the center and $\Sigma(R)$ is the projected stellar density. 

This formulation necessitates the use of a stellar profile. We assume a Plummer profile \cite{1911MNRAS..71..460P}
\begin{equation}\label{eqn:plummer}
\nu(R) = \frac{3L}{4\pi R_e^3} \frac{1}{(1+R^2/R_e^2)^{5/2}} \, ,
\end{equation}
for which the projected stellar distribution takes the form
\begin{equation}\label{eqn:plummerProjected}
\Sigma(R) = \frac{L}{\pi R_e^2} \frac{1}{(1+R^2/R_e^2)^{2}} \, . 
\end{equation}

We use the Bayesian inference tool {\tt MultiNest} \cite{MultiNestDocs} as implemented in the python package {\tt PyMultiNest} \cite{pymultinestDoc}. {\tt MultiNest} operates by sampling $N$ points from the input prior space, then discarding the lowest likelihood $L_0$ point. It is replaced by a new point with likelihood $L_1$ if $L_1>L_0$, and the prior volume is reduced. Going from lowest to highest likelihoods in this way makes it easier to sum up the likelihood over the prior volume later to compute a model's evidence, making this tool well suited to comparing models for selection. 
 
Following the analysis in \citet{GeringerSameth2015}, this is implemented with the unbinned Gaussian likelihood function\footnote{For binned analysis and differences between binned and unbinned analyses \cite{2015MNRAS.453..849B}.}
\begin{equation}\label{eqn:likelihood}
L = \prod_{i=1}^N \frac{1}{\sqrt{2\pi} \sqrt{ \delta_{u,i}^2 + \sigma^2(R_i) } } \exp\left[ -\frac{1}{2} \frac{\left(u_i - \langle u\rangle\right)^2}{  \delta_{u,i}^2 + \sigma^2(R_i)  }
\right].
\end{equation}
Here, $u_i$ is the projected velocity, $R_i$ is the projected position, and $\delta_{u,i}$ is the observational error in velocity of the $i$th star in the data set. $\langle u\rangle$ is the bulk velocity, which is marginalized over with a flat prior. 

%%%%%%%%%%%%%%%%%%%%%%%%%%%%%%%%%%%%%%%%%%%%%%%%%%%%%%%%
%            			Profiles
%%%%%%%%%%%%%%%%%%%%%%%%%%%%%%%%%%%%%%%%%%%%%%%%%%%%%%%%

\section{\label{sec:profiles} Halo profiles}

A key ingredient in using stellar velocities to reconstruct the dark matter distribution in dwarf galaxies is the assumed functional form of the dark matter density profile. 
In order to explore the viability of ultralight bosonic dark matter in dwarf galaxies it is necessary to start from a basic description of the non-linear evolution of halos. This is a difficult problem where the only way to obtain such information is through numerical simulations.

Below we first summarize the distribution of cold dark matter in halos, namely the Navarro Frenk \& White  generalized profile (NFW hereafter) \cite{NFWprofile,1996MNRAS.278..488Z, hernquist1990}. We then describe three different prescriptions of the distribution of dark matter in fuzzy dark matter halos. All three are based on an internal structure that contains a quantum mechanical pressure-supported core. How the core transitions to the outer NFW-like dark matter distribution is the subject of these three models. The differences are summarized in Table~\ref{tab:summary}.

\subsection{Cold dark matter distribution -- NFW profile}\label{subsec:NFWprofile}

The NFW profile \cite{NFWprofile} and subsequently its more generalized form \cite{1996MNRAS.278..488Z, hernquist1990} are the outcome of N-body dark matter simulations where initial thermal velocities in the dark matter are negligible and do not affect the growth of structure (cold dark matter). The form of the dark matter distribution is given by a generalized NFW,
\begin{equation} 
\label{eq:NFW}
\rhoNFW(r) = \frac{\rho_s}{(r/r_s)^\gamma [ 1 + (r/r_s)^\alpha]^{(\beta - \gamma) / \alpha}},
\end{equation}
where  $\rho_s$ and $r_s$ are the characteristic density and scale radius respectively, and $\{\alpha, \beta, \gamma\}$ describe the power law behavior of the dark matter distribution. The profile has an inner density profile that goes as $\sim r^{-\gamma}$ and an outer behavior characterised by $\sim r^{-\beta}$. The normalization of such a profile is specified either by the characteristic density $\rho_s$ and the scale radius $r_s$, or by the mass of the halo $M_{\Delta} = \int \rho(r) d^3r$ and its concentration $c = R_\Delta / r_s$, where $R_\Delta$ is the  radius of the halo. 

One has the freedom to choose how to define a halo, for example whether a halo is defined as a virial overdensity (in this case $\Delta = {\mathrm{vir}}$) or a fixed product of $\Delta$ times the mean matter density of the universe (e.g., $ \Delta =200$). In what follows, when we refer to the mass of an NFW profile we will be using $\Delta =200$, i.e., the NFW profile can be characterized by $\M200$ and $\c200$ (or $R_{200}$). 

This functional form of dark matter distribution has been extensively studied in numerical simulations and has been applied in studies aimed at reconstructing the gravitational potential of dark matter halos on many scales, from galaxy clusters
\cite{2011arXiv1108.5736G}
to the Milky Way \cite{2002ApJ...573..597K} and dwarf galaxies \cite{GeringerSameth2015, Walker2013}. 

When implemented in {\tt MultiNest}, the generalized NFW parameters are sampled over flat priors in linear space for the powers $\alpha, \beta, \gamma$ and in logarithmic space for the parameters $(1-\beta_a)$, $\M200 / M_\odot$, and $\c200$:
\begin{equation}\label{eqn:nfw_priors}
\begin{aligned} 
 -1 \leq -\log_{10} (1-\beta_a) \leq +1, \\
 \log_{10}(5\times 10^7) \leq \log_{10} (\M200 / M_\odot) \leq \log_{10}(5\times 10^9) , \\
 \log_{10}(2) \leq \log_{10} (\c200) \leq \log_{10}(30) , \\
 0.5 \leq \alpha \leq 3, \\
 3\leq \beta \leq 10, \\
 0 \leq \gamma \leq 1.2. \nonumber
\end{aligned}
\end{equation}
Note that the original NFW profile has a power law behavior given by $ \{\alpha, \beta, \gamma\}=\{1,3,1\}$. The priors for $\M200$ have an upper limit at $\M200=5\times 10^{9} \Msun$ because increasing that limit has minimal effect on the posteriors. 

\subsection{Soliton cores}\label{subsec:solcores}
Fuzzy dark matter distribution in collapsed halos is a highly non-linear process that necessitates the use of numerical simulations. The large scale cosmological simulations of \cite{Schive2014} found that axion-like dark matter does lead to the formation of cores that reside in the center of dark matter halos. The density of such cores at $z=0$ (present epoch) is parameterized as 
\begin{equation}\label{eqn:rhosoliton}
\rhosoliton(r) = \frac{1.9 (10 m_{22} )^{-2} (r_c/{\mathrm{kpc}})^{-4}}{ \left[ 1 +9.1\times 10^{-2} \left(r/r_c \right)^2 \right]^8}
10^9 M_\odot \text{kpc}^{-3} ,
\end{equation} 
where $m_{22} \equiv m/10^{-22} $eV is the scaled dark matter particle mass and $r_c$  is the characteristic radius, defined to be the radius at which density drops to one half of the halo's peak value defined as $\rhosoliton( r \rightarrow 0)$. The functional form of Eq.~\ref{eqn:rhosoliton} is accurate to 2\% for $0<r\lesssim3r_c$ \cite{Schive2014}.

The soliton core extends out to the characteristic radius\footnote{The original fitting function from \cite{Schive2014} was in terms of $\Mvir$. Here, for consistency throughout the paper we use the relationship between $\M200$ and $\Mvir$ for a matter density of $\Omega_{\mathrm{M}} = 0.3$ \cite{White2001} to express Eq. \ref{eqn:rc} in terms of $\M200$.},
\begin{equation}\label{eqn:rc}
 r_c \approx 1.5\, m_{22}^{-1} \left( \frac{\M200}{10^9 M_\odot} \right)^{-1/3} \text{kpc}. 
\end{equation} 
For the full wave dark matter density profile of Eq.~\ref{eqn:rhosoliton}, the numerical simulations  of  \citet{Schive2014,Mocz2018} show  that at $\sim 3r_c$ there is a smooth transition to an NFW-like profile.

There is however  ambiguity in how the NFW profile is defined in this case ($\{ \rho_s, r_s$\}, or $\{\M200, \c200\}$) and how it relates to the characteristics of the soliton, namely, $\{\M200,m_{22}\})$ in Eqs.~\ref{eqn:rhosoliton} \& \ref{eqn:rc}. In other words, how is the inner part of the halo (formed early on) related to the distribution of matter in the outskirts of the halo? 

Previous work proposed different methods on how  to make this transition. In this paper we will examine how choices affect the posteriors using stellar kinematics in dwarf galaxies.

\subsubsection{Model A}\label{subsec:GENprofile}

The simplest soliton-like profile is one where the soliton core transitions to an NFW profile at a radius of $\sim 3r_c$ \cite{Safarzadeh_2020}. This is an artificially constructed halo with no physical input other than the characteristics of the soliton core and the outer functional behavior of the NFW profile. There is no imposed physical connection between the two.  

It is composed of two profiles that are matched to have equal densities at the transition radius 
\begin{equation}
\label{eqn:GENcondition}
\rhosoliton\Big|_{r=3r_c} = \rhoNFW\Big|_{r=3r_c} \, .
\end{equation} 

% %           TABLE: Soliton Halo Profile Parameter Constraints

\begin{table*}
\caption{\label{tab:summary} Summary of soliton core models. }
\begin{ruledtabular}
\begin{tabular}{ccL}
% &\multicolumn{2}{c}{$D_{4h}^1$}&\multicolumn{2}{c}{$D_{4h}^5$}\\
 Profile & Free Parameters & Assumptions \\ \hline
  NFW & $\beta_a, \rho_s, r_s, \alpha, \beta, \gamma$ & Cold Dark Matter \\ \hline
  \\
  % GEN : 
  Model A  & $\beta_a, m_{22}, \M200, \c200$ &  Halo mass parameter $\M200$ corresponds to both the soliton mass parameter and the NFW halo mass parameter, transition happens at $3r_c$ with a transition to outer NFW. Total halo mass can be different than $\M200$. \\\hline
  \\ 
  % GM : 
  Model B & $\beta_a m_{22}, \M200', \epsilon$ & Transition happens at $r_\epsilon = f(m_{22}, M_h',...)$, density continuity $\rho_\text{core}(r_\epsilon) = \rho_\text{NFW}(r_\epsilon)$. \\\hline
  \\
  % RBBK : 
  Model C & $\beta_a, m_{22}, \M200$ & Transition happens at $3r_c$, continuous density $\rho_\text{core}(3r_c) = \rho_\text{NFW}(3r_c)$, and second NFW parameter defined by mass conservation.  \\

\end{tabular}
\end{ruledtabular}
\end{table*}

In this model, the free parameters that can define the NFW profile are the mass $\M200$, and concentration $\c200$.  The soliton profile is defined by the same mass $\M200$ and the mass of the scalar dark matter particle $m_{22}$. 

The normalization, $\M200$, and characteristic functional behavior given by $\c200, \alpha, \beta, \gamma$,  of the NFW profile do not need to correspond to physical halos as long as Eq.\ref{eqn:GENcondition} is satisfied.
Here, the NFW mass parameter and the mass that  defines the soliton core is the same, but concentration can vary untethered by the core's form. The model parameters are sampled over the following flat priors:
\begin{equation}\label{eqn:GENsol_priors}
\begin{aligned} 
 -1 \leq -\log_{10} (1-\beta_a) \leq +1, \\
 \log_{10}(5\times 10^7) \leq \log_{10} \left(\M200 / M_\odot\right) \leq \log_{10}(5\times 10^{10}), \\
 \log_{10}(2) \leq \log_{10} (\c200) \leq \log_{10}(120) , \\
 -1 \leq \log_{10}(m_{22}) \leq 3. \nonumber
\end{aligned}
\end{equation}
This model represents the simplest (alas unphysical) prescription for the dark matter distribution in a dwarf galaxy.

\subsubsection{Model B}\label{subsec:GMprofile}

A different approach for connecting the soliton core to the outer parts of the halo was proposed in \citet{GM2017}. Here, the density of the soliton core is fixed to the density of NFW profile at a transition radius that is governed by a free parameter $\epsilon$. In this definition, 

\begin{equation}
    \frac{\rhosoliton}{\left[ 1+ (r_\epsilon/\rsol)^2 \right]^8} = \frac{\rhoNFW}{(1 + r_\epsilon/r_s)^2 (r_\epsilon/r_s)} = \epsilon\rhosol, 
\end{equation}
where
\begin{equation}\label{eqn:repsilon}
    r_\epsilon = \rsol (\epsilon^{-1/8} - 1)^{1/2}
\end{equation}
and $\rsol = r_c / 0.091^{0.5}$ with $r_c$ given by Eq.\ref{eqn:rc} (note that \citet{Schive2014} formulation is equivalent to the formulation by \citet{GM2017} and \citet{MarshPop2015}). 

The density profile in this model is then 
\begin{equation}\label{eqn:GMprofile}
    \rhoGM(r) = \rhosol
    \begin{cases}
      \dfrac{1}{\left[ 1+ (r/r_\text{sol})^2 \right]^8} & r< r_\epsilon \\
      \\
      \dfrac{\delta_\text{NFW}}{(1 + r/r_s)^2 (r/r_s)} & r\geq r_\epsilon       
    \end{cases} 
\end{equation}
where
\begin{equation}\label{eqn:GMdeltaNFW}
    \delta_\text{NFW} = \epsilon \left[ \frac{r_\epsilon}{r_s} \left(1 + \frac{r_\epsilon}{r_s} \right)^2 \right].
\end{equation}

The free parameters chosen by {\tt{MultiNest}} in this model are $\M200$, $\c200$, $m_{22}$, and  $\epsilon$.
These parameters are sampled over the flat priors:
\begin{equation}\label{eqn:GMsol_priors}
\begin{aligned}
 -1 \leq -\log_{10} (1-\beta_a) \leq +1, \\
 \log_{10}(5\times 10^7) \leq \log_{10} \left(\M200 / M_\odot\right) \leq \log_{10}(5\times 10^{10}), \\
 \log_{10}(2) \leq \log_{10} (\c200) \leq \log_{10}(120) , \\
 \log_{10}(0.5) \leq \log_{10}(m_{22}) \leq 3 , \\
 -6 \leq \log_{10}(\epsilon) \leq 1 \nonumber
 \end{aligned}
\end{equation}
Note that for this halo construction it is possible to choose a halo mass parameter that governs core size but results in a different total mass when integrating out to $r_{200}$. 

\begin{figure*}[htbp]
\includegraphics[scale=0.5]{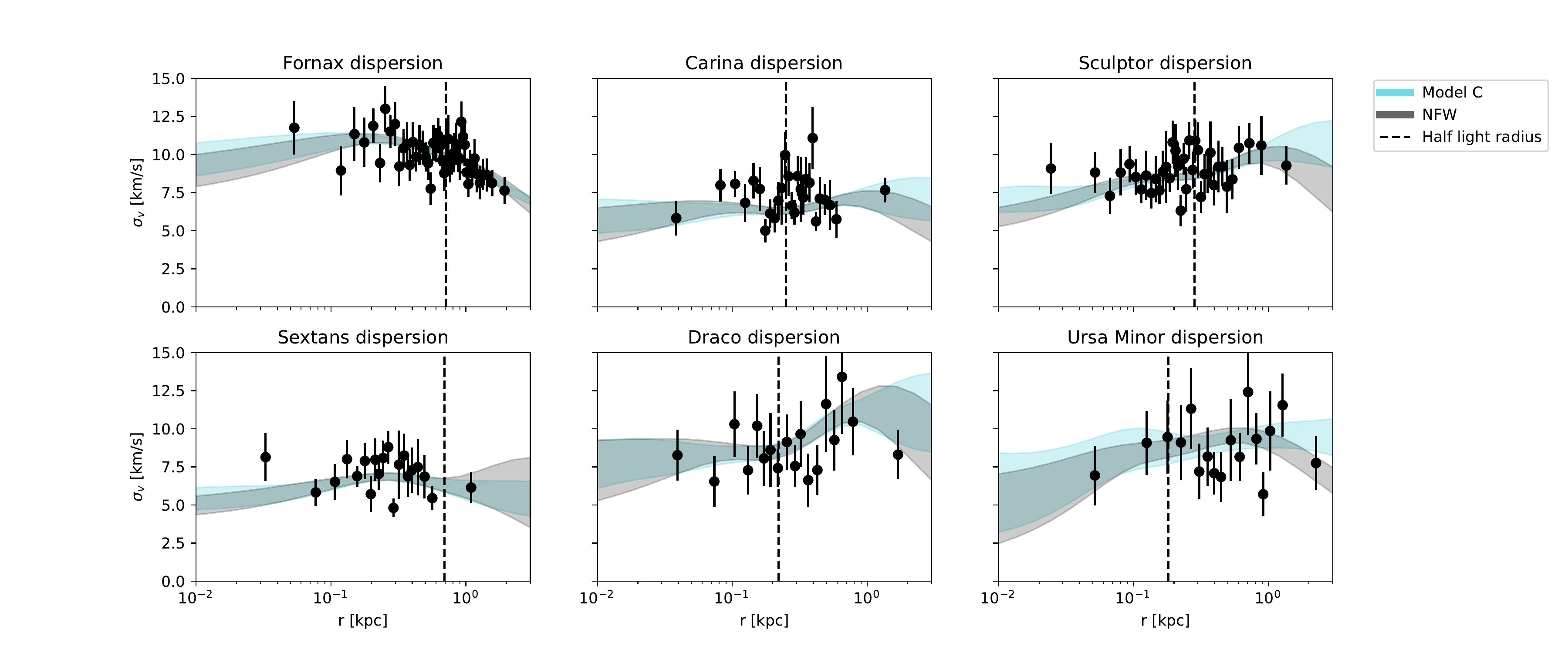}
\caption{Velocity dispersion as a function of radius for the six classical Milky Way Dwarfs. Black points depict binned velocity dispersion measurements (with Poisson error bars). Gray and blue bands represent the 68\% unbinned velocity dispersion from the sampled generalized NFW model and the soliton Model C posteriors, respectively. The vertical dashed line shows the half-light radius. }
\label{fig:fig1}
\end{figure*}

%%%%%%%%%%%%%%%%%%%%%%%%%%%% RBBK

\subsubsection{Model C}\label{subsec:RBBKprofile}

A more physically motivated formulation of the soliton dark matter profile is proposed by \citet{Robles:2018fur}. This formulation connects the inner core of Eq.~\ref{eqn:rhosoliton} to an outer NFW profile at a transition radius $r_\alpha=\alpha r_c$, where $\alpha$ is found to be $\alpha \approx 3$ (see \citet{Schive2014, Mocz2018}). 
\begin{equation}\label{eqn:RBBKprofile}
    \rho_\text{RBBK}(r) = \begin{cases}
      \rho_\text{sol}(r) & 0 \leq r \leq r_\alpha \\
      \rho_\text{NFW}(r) & r_\alpha \leq r \leq r_{200}.
    \end{cases}
\end{equation}
In this model, mass is conserved, and the total mass of a halo is the sum of the mass in the soliton core and the mass of the corresponding NFW profile. In other words,

\begin{equation}\label{eqn:RBBK_masscont}
\M200 = M_\text{core} + 4\pi\int_{r_\alpha}^{r_\Delta} \rho_\text{NFW}(r') r'^2 dr'
\end{equation}

Note that we take $\alpha=3$ as before, although this may vary with $\M200$ as discussed in \cite{Robles:2018fur}. 
One important feature of this profile is that not every combination of $\{\M200, \m22\}$ parameters is valid. This reflects the fact that there is a minimum halo mass set by the core size, which is determined by $\m22$ (small halos are not allowed to form because of quantum pressure).

We implement this model in 
{\tt MultiNest}  by sampling over the free parameters with the following flat priors
\begin{equation}\label{eqn:RBBKsol_priors}
\begin{aligned}
 -1 \leq -\log_{10} (1-\beta_a) \leq +1,  \\
 -1 \leq \log_{10}(m_{22}) \leq 3 \nonumber \\
 \M200^\mathrm{min}(m_{22})/\Msun \leq \log_{10} \left(\M200 / \Msun \right) \leq \log_{10}(5\times 10^{10}),  
 \end{aligned}
\end{equation}
where $\M200^\mathrm{min}(m_{22})$ is obtained from solving the following equation for minimum possible $\M200$ for a given $m_{22}$, 
\begin{equation}\label{eqn:FMmin}
\begin{aligned}
\frac{{\M200^\mathrm{min}} - M_\mathrm{soliton}\Big|_{r_\alpha}}{4\pi r_\alpha^3 \,\,\rhosoliton\Big|_{r_\alpha}} 
=
\left(1 + \frac{r_\alpha}{r_s} \right)^2 \left( \frac{r_s}{r_\alpha}\right)^2 \\ 
\times
\left[ \frac{r_s(r_\alpha - r_\Delta)}{(r_\Delta + r_s)(r_\alpha + r_s)} 
+ \log \left(\frac{r_\Delta + r_s}{r_\alpha + r_s} \right)
\right] ,
\end{aligned}
\end{equation}
where $r_\alpha = \alpha r_c$, and $r_c$ is given by Eq.~\ref{eqn:rc}, $\rhosoliton$  is given by Eq.~\ref{eqn:rhosoliton}, and $M_\mathrm{soliton}$ is
\begin{equation}
M_\mathrm{soliton} = \int_0^{r_\alpha} \rhosoliton d^3 r, 
\end{equation}
all of these functions of $\M200$ and $\m22$.

%%%%%%%%%%%%%%%%%%%%%%%%%%%%%%%%%%%%%%%%%%%%%%%%%%%%%%%%
%            			DATA     
%%%%%%%%%%%%%%%%%%%%%%%%%%%%%%%%%%%%%%%%%%%%%%%%%%%%%%%%
\section{\label{sec:data} Observations}
For the dwarf galaxies Carina, Fornax, Sculptor and Sextans  we adopt the stellar-kinematic data sets of  \citet{WalkerMateoOlszewski2009}.  We refer the reader to \citet{2007ApJS..171..389W} for a detailed description of the target selection, observation, and data reduction methods.  In order to identify stars that are members of each dwarf galaxy (as opposed to foreground contaminants contributed by the Milky Way), we adopt the membership probabilities estimated by \citet{2009AJ....137.3109W}, which are derived under the simplifying assumption that the velocity dispersion within each system is constant with projected galactocentric distance.  Selecting only those stars having membership probability $>95\%$, the samples contain 774, 2483, 1365 and 441 probable members of Carina, Fornax, Sculptor and Sextans, respectively.  

For  Draco and Ursa Minor, we adopt stellar-kinematic data sets of \citet{Spencer_2018}, who include catalogs from multiple literature sources spanning 30 years of observations.  Applying the same simple model for distinguishing member stars from foreground contaminants, we obtain a data set containing 692 probable member stars for Draco (341 stars which have multiple observations, that are combined into a single measurement by taking the mean velocity, weighted by the inverse-square of the velocity error) and 680 for Ursa Minor (284 stars that have multiple observations).  

For all galaxies, we adopt the half light radii originally published by  \citet{1995MNRAS.277.1354I}. 

%									FIGURE

%								
\begin{figure*}[htbp]
\includegraphics[scale=0.5]{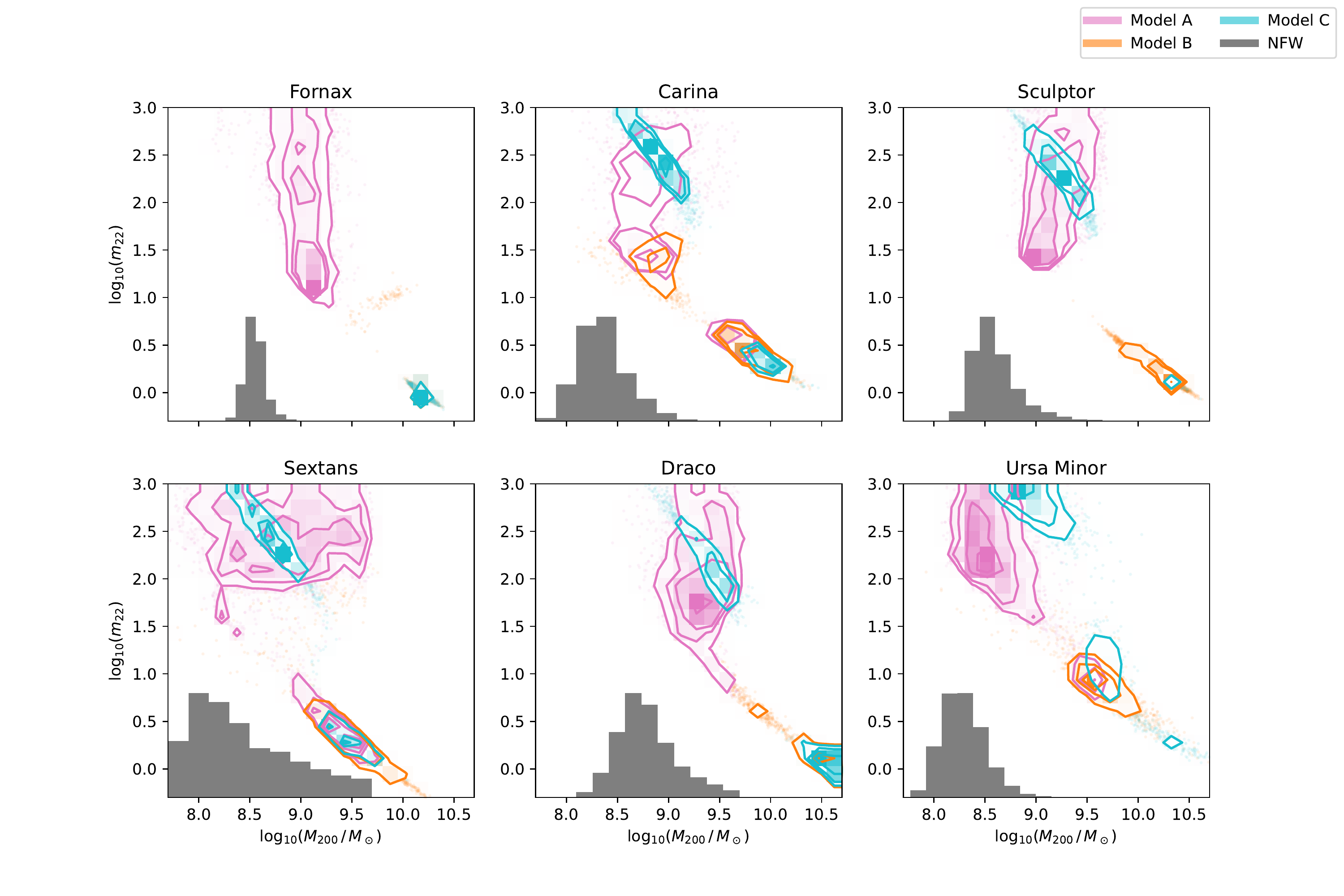}%
\caption{Halo mass and $m_{22}$ posteriors for the six Milky Way dwarf galaxies. Pink contours correspond to Model A, orange contours correspond to Model B, and blue contours represent Model C. The gray histogram represents the halo mass posteriors for a generalized NFW profile. Note that Model B's contours in Fornax lie directly under Model C's. This illustrates the anticorrelation between $\m22$ and $\M200$ -- high values of $m_{22}$ require  low halo masses, and low values of $m_{22}$ require high halo masses. This result is due to changes in the velocity dispersion anisotropy -- see text and Fig.~\ref{fig:fig3} for details.  }
\label{fig:fig2}
\end{figure*}

%%%%%%%%%%%%%%%%%%%%%%%%%%%%%%%%%%%%%%%%%%%%%%%%%%%%%%%%
%            		 RESULTS  
%%%%%%%%%%%%%%%%%%%%%%%%%%%%%%%%%%%%%%%%%%%%%%%%%%%%%%%%

\section{\label{sec:results} Results}

In Figure~\ref{fig:fig1} we show an example of fits to the velocity dispersion data for all six dwarf galaxies. Note that the data shown in Fig.~\ref{fig:fig1} is binned (for illustration purposes), but the fits are obtained from the unbinned analysis as described in Section~\ref{sec:SecII}). We compare the posterior distributions in velocity dispersion of the generalized NFW profile with one of the soliton core profiles, Model C \cite{Robles:2018fur}. The reason we choose Model C is because it is the most physically motivated description, and allows a more direct comparison with the NFW profile as mass is conserved in both cases (the sum of soliton core mass and the mass distributed as NFW in Model C is the same as the total mass of an NFW-only profile). The fits show how the data is most restrictive at the half-light radius, with the outskirts of the halos more unconstrained as expected\footnote{The other two soliton prescriptions (Model A and Model B \cite{GM2017}) have similar behavior in that regard.}.

Figure~\ref{fig:fig2} shows the main result of this paper. For all dwarf galaxies we find that data allows two distinct regions in the $\M200 - \m22$ parameter space. One requires a low value of $\m22$ and high $\M200$, while the other is the opposite, i.e., high values of $\m22$ and low $\M200$.  The reason is because as halo mass increases, the size of the soliton core gets smaller as $r_c \sim \M200^{-1/3}$ and the density is higher as $r_c \sim \M200^{4/3}$. In other words, as particle mass increases, quantum effects will become less pronounced and thus the dark matter distribution behaves more classically, and more NFW-like.

This behavior may be related to the same degeneracy that exists between $\rho_s$ and $r_s$ when one fits NFW and/or generalized NFW profiles to dwarf galaxies \cite{2006ApJ...652..306S,2009ApJ...704.1274W,2008Natur.454.1096S}. The origin of such degeneracy is the fact that the Jeans equation is most constraining at the Plummer radius \cite{2008Natur.454.1096S}. If all models are then forced to have same mass interior to the Plummer radius, then it is possible to have models with anticorrelated $\rho_s$ and $r_s$ being equally good fits to the data. 
\begin{figure*}[htbp]
\includegraphics[scale=0.6]{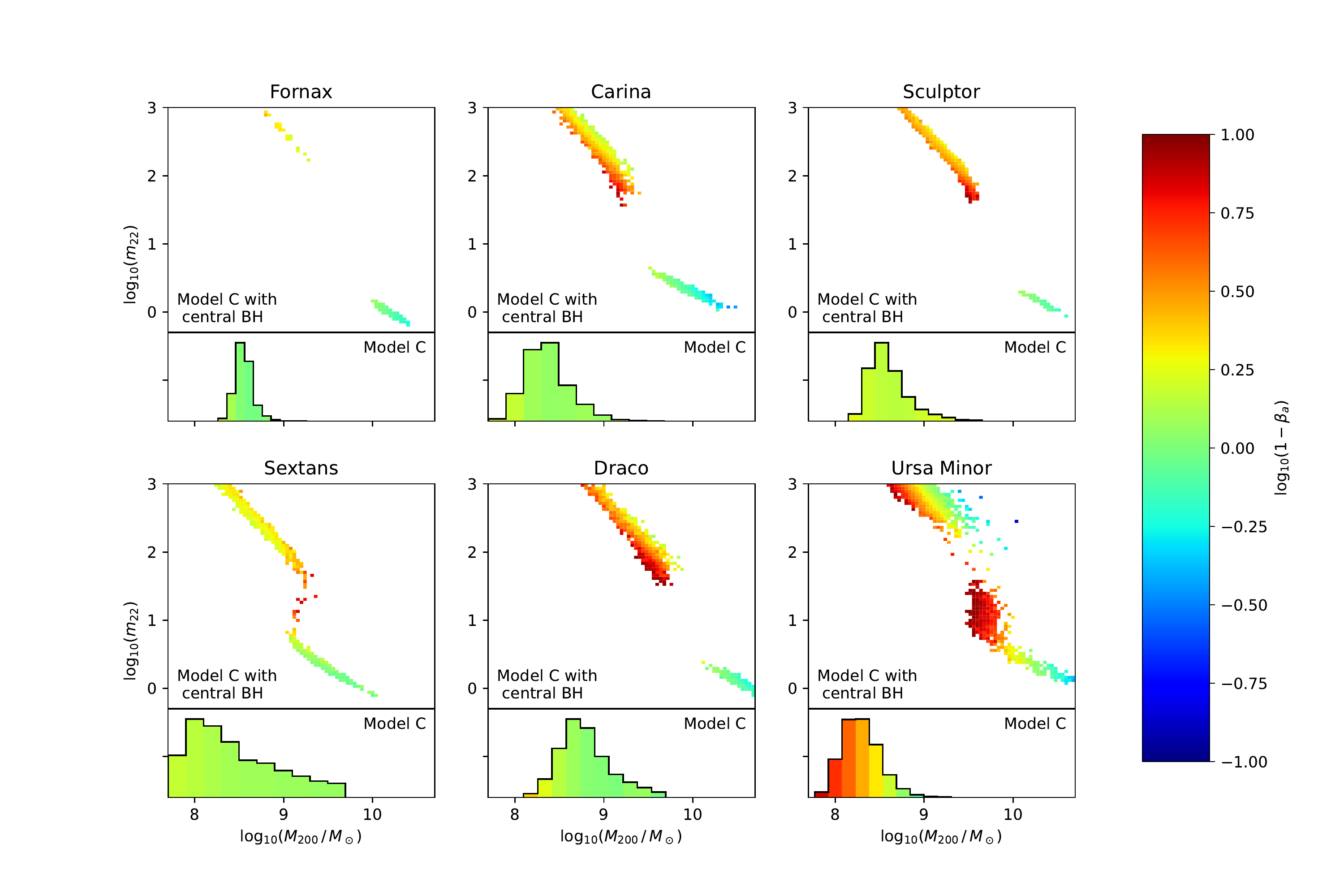}
\caption{Same as Fig.~\ref{fig:fig2} but as a posterior scatter plot in the $m_{22}- \M200$ parameter space for Model C in all six dwarfs. Color corresponds to the anisotropy parameter. High mass halos that favor $\m22 \sim 0$ are more radially biased compared to lower mass halos that favor $\m22 \gtrsim 2$. The histogram corresponds to the mass posterior of an NFW fit to the data, with color corresponding to the average velocity anisotropy in each bin of the histogram.}
\label{fig:fig3} 
\end{figure*}

There are a few ways to understand why these different regions of the $\M200 - \m22$ parameter space are allowed by the data. 
One way is to consider the concentration in the distribution of dark matter; a cusped profile will have more mass concentrated in the center as compared to a cored profile of the same total mass, so the total mass for a cored model must increase in order to fit the inner data points. 

Another way to understand what drives the fit is to look at the anisotropy (see Section~\ref{sec:SecII} as to the physical reasoning of how anisotropy can affect the fit). 
In Fig.~\ref{fig:fig3} we show the $\m22, \M200$, and anisotropy posteriors for Model C (shown as a colored scatter plot), as well as $\M200$ and anisotropy posteriors for the NFW model (shown as a histogram). It is clear that Model C soliton posteriors tend to have a higher $\log_{10}(1-\beta_a)$ than NFW\footnote{Models A and B have similar behaviour.}. A higher $\log_{10}(1-\beta_a)$ means a more tangentially biased halo with lower velocity anisotropy, which causes a suppressed velocity dispersion at low radii. This subtle effect allows models with similar halo mass but different mass distribution and velocity dispersion in the inner radii to both be consistent with data.  

A notable feature in Figure \ref{fig:fig3} is a bimodal effect for Model C posteriors, where in addition to high $\log_{10}(1-\beta_a)$ there is a mode with lower $\log_{10}(1-\beta_a)$. For models with $\log_{10}(1-\beta_a)\sim 0$, the velocity dispersion is neither amplified or suppressed. For models with a negative $\log_{10}(1-\beta_a)$, this corresponds to more radially biased halos (higher $\beta_a$), and an amplified velocity dispersion at low radii \cite{GalacticDynamics}. As shown in Figure \ref{fig:fig3}, these modes also correspond to low $m_{22}$ and high halo mass. As a result, these posteriors fit the velocity dispersion at low radii well, either because they are cored profiles with high density or the radial anisotropy is radially amplified\footnote{This effect is more pronounced in dwarf galaxies without stars at very low radii. For example in Draco, the inner most star in the unbinned data is $\sim 20$ pc, as compared to Sculptor with an inner most star at $\sim 5$ pc.}.

Note that if we restrict the maximum limit in the prior of the particle mass to $\log_{10}\m22 \leq 1.5$, we find similar results as in \citet{GM2017}, namely a degeneracy between anisotropy and $\m22$ (or core size) -- lower $\m22$ values are preferred when the anisotropy is allowed to be a varying constant instead of fixed at $\beta=0$. However,  when increasing the maximum allowed particle mass to $\log_{10}\m22 \leq 3$, higher particle masses (and lower core sizes) are also allowed when anisotropy is a varying constant. Therefore, it is not clear that low $\m22$ values are a physical outcome of the prior in the anisotropy \cite{GM2017}, and instead this effect may be the result of a restrictive prior in $\m22$.

\begin{figure*}[htbp]
\includegraphics[scale=0.6]{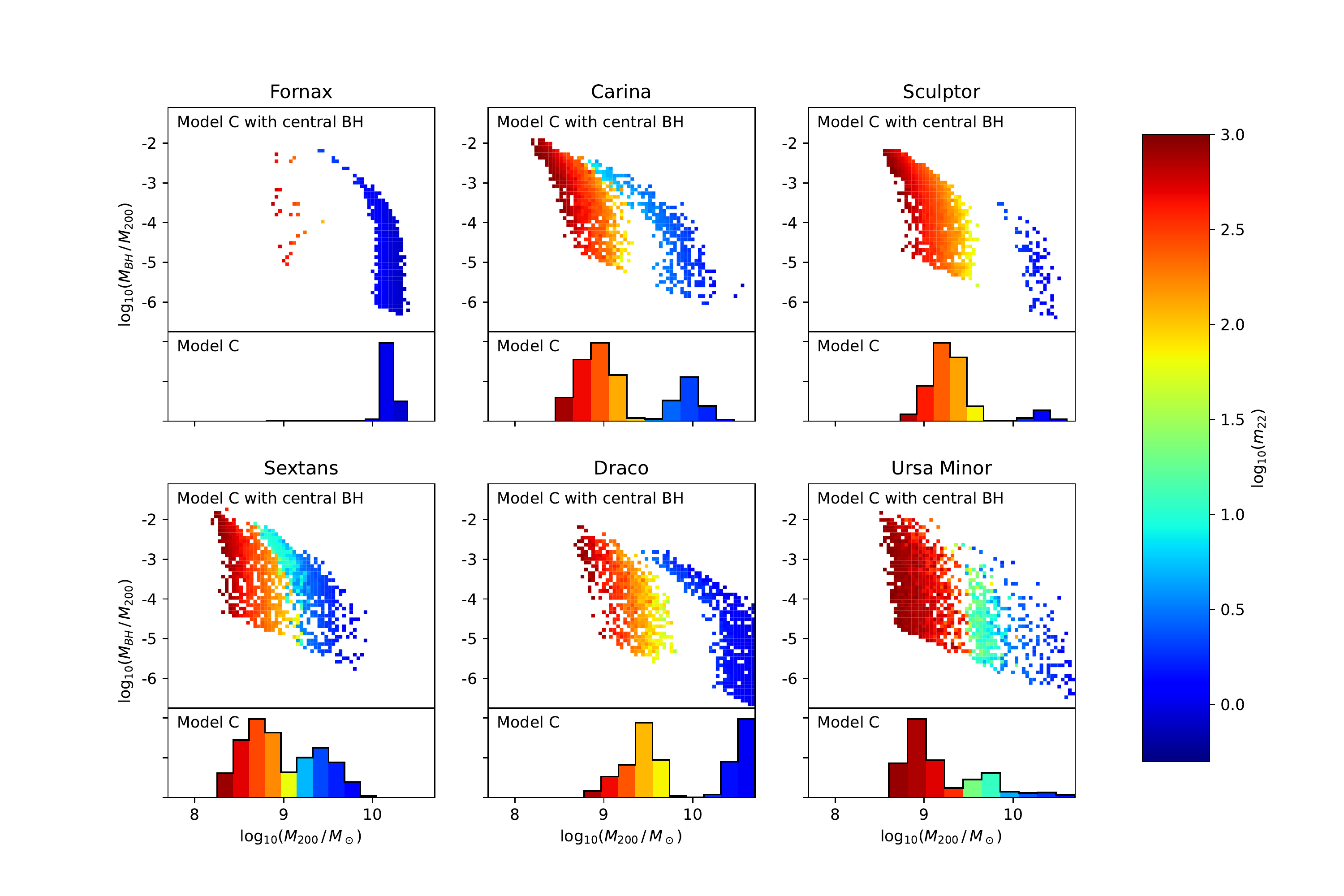}
\caption{Scatter plot of the central black hole posteriors for Model C in the  $\MBH/\M200$ -- $\M200$ parameter space. Color corresponds to the value of $\m22$. For comparison, the histogram depicts  $\M200$ posterior for Model C without a black hole. Histogram color represents the average value of $\m22$ in a given bin. }
\label{fig:fig4} 
\end{figure*}

The degeneracy in the $\m22-\M200$ plane raises a question as to the meaning of halo mass. We can interpret $\M200$ as the mass of the halo in the field and before its interaction with the Milky Way. In other words, the mass of the halo today must be less than the mass quoted above. If $\m22 \sim 0$, then all six dwarfs require that the mass of their host dark matter halo is approximately $\M200 \sim 10^{9.5-10}\Msun$. These halo masses are large compared to the masses obtained with NFW fits (see gray histogram in Fig.~\ref{fig:fig2}). We already know that the most massive dwarfs in the Milky Way are the Large and Small Magellanic Clouds, which means that the six dwarfs as implied here ($\m22 \sim 0$) are of order ${\cal{O}} \sim 10^{-1}$ the mass of the Large Magellanic Cloud \cite{10.1093/mnras/stz1371}.

The question then is whether it is possible for the Milky Way to host at least six dwarf galaxies with masses of order ${\cal{O}} \sim 10^{10}\Msun$. We can estimate this probability  of a dwarf galaxy having a mass of order ${\cal{O}} \sim 10^{10} \Msun$ by estimating the probability that such halo was formed at some time in the past and it is now part of the Milky Way.

Linear perturbation theory allows such estimate. If a perturbation crosses a critical overdensity threshold, $\delta_{\rm{collapse}} \approx 1.686$, then the overdensity will virialize and form a halo of mass $M$ at redshift $z$. The rareness of this fluctuation is encapsulated in the standard deviation of fluctuations that contain a mass $M$ at redshift $z$. 
\begin{equation} 
\label{eqn:variance}
\nu(\M200,z) = \frac{\delta_{\rm{collapse}}}{\sigma(\M200) {\cal{D}}(z)},
\end{equation} 
where $\sigma(\M200)$ is the variance on scale $\M200$ and ${\cal{D}}(z)$ is the growth factor of linear perturbations, defined such that ${\cal{D}}(z=0)=1$ \cite{2003moco.book.....D}. 

The probability that such object merged with the Milky Way is then obtained by 
\begin{equation} 
\label{eqn:EPSprobability}
{\cal{P}} (\M200, z) = 1 - \Phi [\nu(\M200,z)]^{{\MMW/\M200}}.
\end{equation} 
This formulation accounts for the fact that there are $\MMW/ \M200$ distinct halos that {\it{could}} have merged to form the Milky Way (trials factors). The function $\Phi[\nu(\M200,z)]$ is the cumulative distribution function of a standard normal distribution (with mean 0 and variance 1), and $\nu(\M200,z)$ given by Eq.~\ref{eqn:variance}.

If we assume $\MMW \approx 10^{12} \Msun$ and $\M200 \approx 10^{10} \Msun$ then $\MMW/ \M200 \approx 100$, therefore the probability of the Milky Way to have merged with one $10^{10}\Msun$ object is $P \approx 0.15$ (using $\Omega_M = 0.3$, $\Omega_\Lambda = 0.7$, and $h=0.7$ in Eq.~\ref{eqn:variance}). The probability that all six of the dwarfs we study here have originated from a $10^{10}\Msun$ halo is thus $0.15^6  \approx 10^{-6}$. We therefore conclude that {\it either $\m22$ must be greater than $\m22 \sim 1$, or that the Milky Way halo is not a typical $10^{12}\Msun$ halo.}

There is however one other possibility for halos to be described with a low $\m22$ {\it and} a low $\M200$. And that is the presence of a  massive black hole embedded in a soliton core. The physical mechanism that can lead to such objects in dwarf galaxies is highly speculative, but nevertheless it has been considered  as a mechanism for generating cored profiles within standard cold dark matter \cite{Silk_2017}. Such limits have been placed in field dwarf galaxies, as in \citet{Reines2011}. 

Here, we can use the same formalism described in Sections \ref{subsec:NFWprofile} and \ref{subsec:RBBKprofile} to ask the question whether a Milky Way dwarf galaxy with a soliton core can mimic the velocity dispersion profile of a cusped profile (either described by an NFW profile or by bosonic dark matter with an $\m22 \sim 10$ or greater. 

We can implement this in the {\tt{MultiNest}} analysis with the addition of a point mass at the center of each dwarf. In both cases, the mass of the black hole is sampled over a flat prior:
\begin{equation}\label{eqn:mBH_prior}
\begin{aligned} 
 4 \leq \log_{10}(\MBH / M_\odot) \leq 9 \nonumber
\end{aligned}
\end{equation}

Figure \ref{fig:fig4} shows a comparison of $\m22, \M200$, and $\MBH$ posteriors for Model C with a central black hole (shown as a colored scatter plot) to $\m22$ and $\M200$ posteriors for Model C without a black hole (shown as colored histograms). What is observed at low black hole mass is the two distinct $\m22-\M200$ regions for low black hole masses. However, as the black hole mass increases, models with low $\m22$ and high $\M200$ now prefer halos of lower mass. A halo with a black hole of order ${\cal{O}} \gtrsim 10^{-3}$ of the halo mass can have a soliton core with low $\m22$. In other words, by adding a central point mass to a cored profile, it is possible to mimic the effects of a cusped profile (for a similar result but a different analysis in Leo 1 see  \citet{2021ApJ...921..107B}).

Given these different models it is illuminating to ask the question whether one model is preferred over another. This can be obtained using the log-likelihood ratio, simply the ratio of the likelihood of one model to the likelihood of another. {\tt MultiNest} is particularly well suited to comparing models; it works by keeping a set of live points sorted by their likelihood, and replacing the lowest point only if the next point drawn has a higher likelihood.

\begin{figure}[htbp]
\includegraphics[scale=0.6]{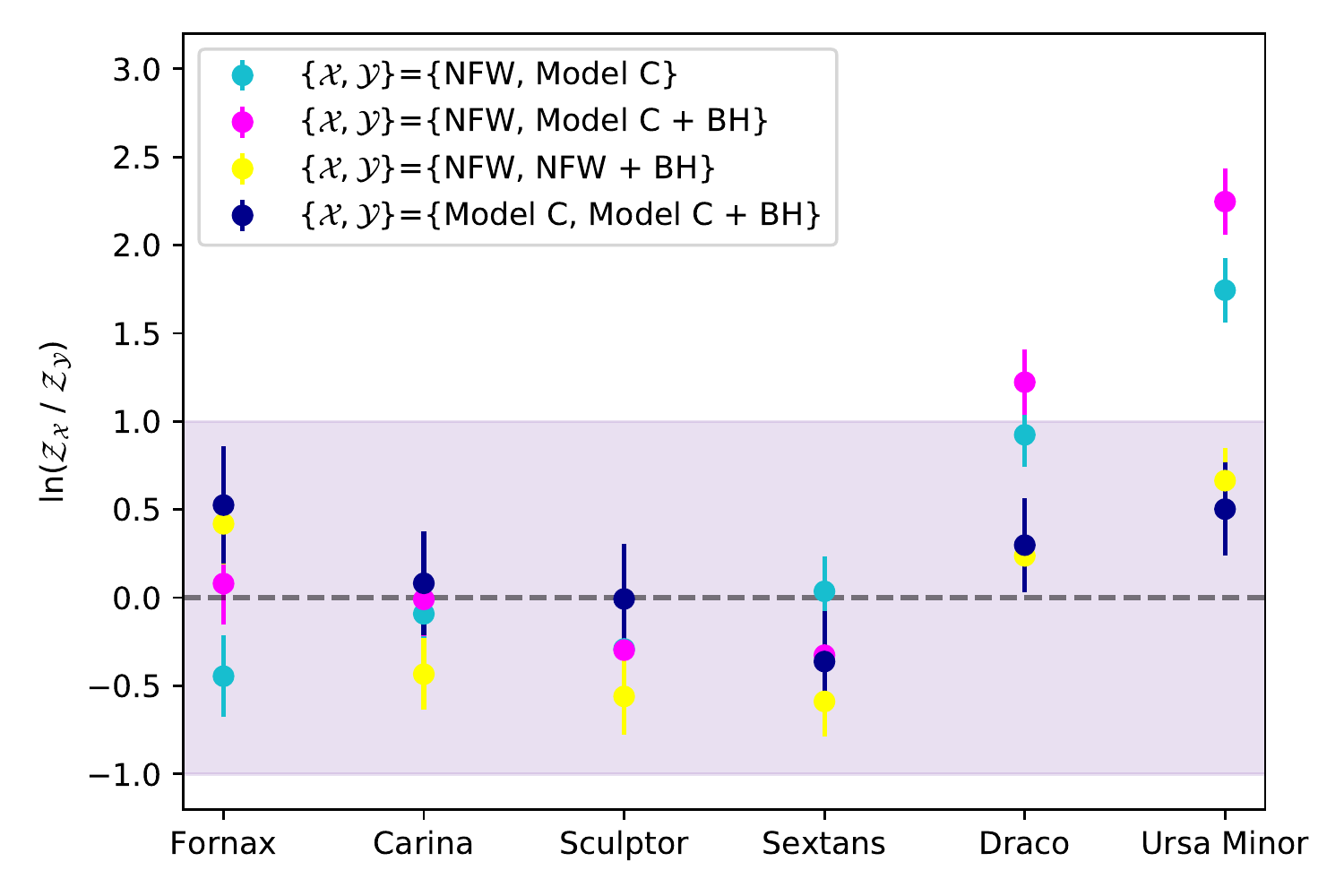}%
\caption{Model comparison for the six dwarf galaxies shown as the logarithm of the evidence ratio $\ln(\mathcal{Z}_X /\mathcal{Z}_Y)$. Positive values favor model X, and negative values  favor model Y, where X and Y correspond to different models as shown in the legend. Values greater than one, outside of the purple band, are generally considered good evidence. Note that Draco and Ursa Minor have the least number of stars in this sample -- see text for details.}
\label{fig:fig5}
\end{figure}

The evidence is the sum of likelihood over the prior volume, which can be calculated efficiently from the live points after convergence. 
In Figure~\ref{fig:fig5} we show the log likelihood ratio $\log{(\mathcal{Z}_X / \mathcal{Z}_Y)}$ where $X$ and $Y$ are the two models being compared. 
A log-likelihood ratio of greater than 10 is generally considered to be good evidence preferring one model over the other \cite{Jeffreys1961}, represented as being outside the shaded purple band. 
Positive values prefer Model $X$ and negative prefer Model $Y$. 

There is no  evidence of one model being preferred over another in most dwarfs, with the exception of Ursa Minor preferring an NFW to the Model C  with or without a central black hole. However, Ursa Minor is the dimmest and most irregular of the considered galaxies, with the fewest number of stars, as well as evidence of tidal disruption and so should be not be taken at face value \cite{2001ApJ...549L..63M, 10.1093/mnras/stu938}. 
Note also that even though it is possible to have a central black hole and reasonably low mass halo with low $\m22$, the evidence does not favor the model with a black hole to the model without a black hole.  

%									FIGURE

%%%%%%%%%%%%%%%%%%%%%%%%%%%%%%%%%%%%%%%%%%%%%%%%%%%%%%%%
%            		 CONCLUSION   
%%%%%%%%%%%%%%%%%%%%%%%%%%%%%%%%%%%%%%%%%%%%%%%%%%%%%%%%

\section{\label{sec:conclusion} Conclusion}

In this paper we explore the viability of ultralight bosons as the dark matter in dwarf galaxies. This is motivated by the large scale properties of the distribution of such dark matter candidates. The formation of a soliton core (on kpc scales) due to quantum pressure has been proposed as a solution to the core/cusp problem in dwarf galaxies  \cite{WalkerPenarrubia2011, Pascale2018, Hayashi2020, ReadWalkerSteger2019}, and other small scale issues in galaxy formation \cite{Schutz2020}.

We use stellar velocity dispersion measurements in six classical Milky Way dwarf galaxies, and employ a Jeans analysis to reconstruct the gravitational potential. 
The form of the soliton core is a fit to simulations \cite{Mocz2018, Schive2014}  that depends on the boson mass and the halo mass. In the inner parts quantum pressure sets a core which smoothly transitions to an NFW-like profile in the outer parts of the halo. We consider four different implementations of the core to NFW-like transition. 

We find a multimodal posterior distribution: two distinct anticorrelated regions of particle mass and halo mass. The resulting posteriors show that that there are two allowed regions of the parameter space: low particle masses ($\m22 \sim 0$) along with high halo masses ($\M200 \sim 10^{10}$), or high particle masses ($\m22 \gtrsim 2$, i.e., CDM-like) with lower halo masses ($\M200 \sim [10^8 - 10^9]\Msun$), consistent with \cite{Hui2017}. This is understood in the context of the velocity anisotropy as shown in Figure \ref{fig:fig3},  which can suppress or supplement the velocity dispersion to allow for two regions in parameter space. However taking into consideration the hierarchical merging history of the Milky Way, it is very improbable for a Milky Way size halo to have six ${\cal{O}} \sim 10^{10} \Msun$ subhalos in addition to the Small and Large Magellanic Clouds. Thus the high mass halos required to have low particle masses are very unlikely. 

An alternate viable option for a soliton core to exist in dwarf galaxies is if a black hole is present in the center of the dwarf galaxy. As shown in Figure \ref{fig:fig4}, it is possible to have low particle mass with ${\cal{O}} \sim [10^8 - 10^9]\Msun$ halos with the inclusion of a central black hole with mass $M_{BH} \sim 10^{-2} - 10^{-3} \M200$. This is proportionally a massive black hole in comparison to halo size, especially in context of the black hole and host spheroid mass relationship observed in previous studies (see e.g., \cite{2004MNRAS.354..292K}). Furthermore, no reliable mechanism through galaxy formation or hierarchical structure formation is known to explain their presence. 

Given these models, it is natural then to ask the question whether  any of the models is considered favored by the data.  Figure \ref{fig:fig5} depicts the evidence, a measure of favorability among any two models. We find that this analysis and with the current state of data there is no appreciable difference between an ultralight bosonic dark matter distribution over cold dark matter, nor does it favor a model with a central black hole over a model without. This holds for all of the classical dwarfs considered, with the exception of Ursa Minor (the most irregular of the considered galaxies, and the one with the least amount of stellar velocity dispersion data).

This work is limited by the assumption that anisotropy is constant for a given system, as opposed to letting it vary with radius. This assumption can be relaxed in two ways: first, one can repeat the aforementioned calculation by allowing anisotropy to vary freely. Alternatively, it may be possible to obtain tangential velocities in the near future. If this observational  challenge is accomplished then it will be possible to fully reconstruct the three-dimensional potential without ambiguities arising from assumptions regarding tangential velocities. We plan to address both of these challenging topics in future work.

In summary, we  conclude that ultralight bosonic dark matter of mass $m\lesssim 10^{-20}$eV is extremely unlikely in six of the classical Milky Way dwarf galaxies, unless the Milky Way has a very unusual merger history or each dwarf contains a proportionally massive black hole. In lack of evidence for both of these requirements, we constrain the mass of the dark matter particle to be $m\gtrsim 10^{-20} \mathrm{eV}$.  

We acknowledge useful conversations with Stephon Alexander, Jatan Buch, Steven Clark, Tatsuya Daniel, Ian Dell'Antonio, JiJi Fan, Leah Jenks, John Leung, Alexis Ortega, Michael Toomey, and Kyriakos Vattis. I. S. G gratefully acknowledges a Fellowship from the NASA Rhode Island Space Grant (NASA-80NSSC20M0053) that enabled working on this research.  A faculty research seed grant to S. M. K. from the NASA-Rhode Island EPSCoR Research Infrastructure Development program (NASA NNX16AR01A) also made this study possible.
S. M. K. is partially supported by NSF PHY- 2014052. M.G.W. acknowledges support from NSF grants AST-1813881 and AST-1909584.  
This research has made use of the VizieR catalogue access tool \cite{vizier2000} and 
was conducted using computational/visualization resources and services at the Center for Computation and Visualization, Brown University.

%%%%%%%%%%%%%%%%%%%%%%%%%%%%%%%%%%%%%%%%%%%%%%%%%%%%%%%%%%%%%
%	   								        Bibliography     
%%%%%%%%%%%%%%%%%%%%%%%%%%%%%%%%%%%%%%%%%%%%%%%%%%%%%%%%%%%%%
\bibliography{DwarfsSolitons.bib}

\end{document}